\documentclass[%
reprint,
amsmath,amssymb,
aps,
]{revtex4-2}
\usepackage{graphics}      
\usepackage{graphicx}      
\usepackage{longtable}     
\usepackage{url}           
\usepackage{bm}            
\usepackage[usenames]{color}
\usepackage[dvipsnames]{xcolor}
\usepackage{cancel}
\usepackage{soul}
\usepackage{amsmath,amssymb}
\usepackage{epstopdf}
\usepackage{mathtools}
\usepackage{natbib}
\usepackage{amsfonts}
\usepackage{amsthm,mathrsfs,amsopn}

\theoremstyle{plain}

\begin{document}
\preprint{APS/123-QED}
\title{On the role of zealots in a best-of-n problem on a heterogeneous network}
%
%

%
%
%

\author{Thierry Njougouo}
\affiliation{Namur Institute for Complex Systems (naXys), University of Namur, Namur, Belgium}

\author{Andreagiovanni Reina}
\affiliation{IRIDIA, Université Libre de Bruxelles, Brussels, Belgium}

\author{Elio Tuci}
\affiliation{Namur Institute for Complex Systems (naXys), University of Namur, Namur, Belgium}

\author{Timoteo Carletti}
\affiliation{Namur Institute for Complex Systems (naXys), University of Namur, Namur, Belgium}
\affiliation{Department of Mathematics, University of Namur, Namur, Belgium}


\date{\today}

\begin{abstract}
Both humans and social animals live in groups and are frequently faced to choose between options with different qualities. When there are no leader agents controlling the group decision, consensus can be achieved through repeated interactions among group members. Various studies on collective decision-making illustrate how the dynamics of the opinions are determined by the structure of the social network and the methods that individuals use to share and update their opinion upon a social interaction.
In this paper, we are interested in further exploring how cognitive, social, and environmental factors interactively contribute to determining the outcome of a collective best-of-$n$ decision process involving asymmetric options, i.e., different costs and/or benefits for each option.
We propose and study a novel model capturing those different factors, i) the error in processing social information, ii) the number of zealots (i.e., asocial agents who never change their opinion), iii) the option qualities, iv) the social connectivity structure, and v) the degree centrality of the asocial agents. By using the heterogeneous mean-field approach, we study the impact of the above-mentioned factors in the decision dynamics. Our findings indicate that when susceptible agents, i.e., individuals who change their opinion to conform with others, use the voter model as a mechanism to update their opinion, both the number and the degree of connectivity of the zealots can lead the population to converge towards the lowest quality option. Instead, when susceptible agents use methods more cognitively demanding, the group is marginally impacted by the presence of zealots.
The results of the analytical model are complemented and extended by agent-based simulations. 
Our analysis also shows that the network topology can modulate the influence of zealots on group dynamics. In fact, in homogeneous networks where all nodes have the same degree (numbers of neighbours), any location of the zealots has similar impact on the group dynamics. Instead, when the network is heterogeneous, our simulations confirm the model predictions that show that placing the zealots in the network hubs (nodes with several neighbours) has a much larger impact than placing them in lower-degree nodes.
\end{abstract}

\maketitle
\section{Introduction}
\label{sec::intro}
Many decisions characterising the social life of humans as well as of other social species are the consequence of information exchange among several individuals and  can be considered to be complex processes. In spite of the fact that the outcomes of these decisions influence the life of all individuals of a community, in some cases the decision process is governed by a limited number of dominant individuals (e.g., political leaders in humans, or socially dominant individuals in primates). In other cases, the decisions are genuinely collective, meaning that they are the results of social interactions between peers, and are characterised by the fact that the decision made is not attributable to any individual of the group \citep{conradt2005consensus}. For instance, primate groups like chimpanzees and baboons engage in a collective decision-making process to determine their direction after a period of rest \citep{strandburg2015shared, sueur2011group, wang2020decision}. Similarly, a flock of birds collectively decides when to leave a foraging patch~\citep{farine2014collective, bidari2022stochastic}
a swarm of honeybees, during reproductive swarming, decides on the site where to build their new nest ~\cite{beekman2018different,seeley2001nest,seeley2004group,reina2017model}.

Scenarios requiring collective decisions have been studied in different scientific disciplines and with different methods, such as experimental methods~\citep{conradt2005consensus,conradt2009group,goeree2011experimental}, computational modelling and simulations methods~\citep{lambiotte2008majority,beal2021intra,mann2018collective}
and social network analysis~\citep{centola2018behavior,sueur2012social,siegel2009social}. 
Other study~\citep{valentini2017best} provided a descriptive framework for collective decision-making processes based on a costs/benefits analysis for sampling and selecting the available options. In particular, they distinguished between symmetric conditions (i.e., same costs and/or same benefits for each option) and asymmetric conditions (i.e., different costs and/or different benefits for each option). A series of work have shown that consensus for the best alternative can be achieved even when individuals operate without a leader and make noisy individual estimates of the option's qualities (benefits/costs). Indeed, to make collective best-of-n decisions, it is sufficient that the individuals use the estimated quality to modulate the frequency (or persistence) with which they share their opinion. This mechanisms has been observed in group-living animals, e.g., ants and honeybees \cite{seeley2001nest}, and then employed to design artificial distributed systems, such as robot swarms \cite{Parker2004, Reina2015}. There are multiple causal factors, interacting in a complex way, that contribute to determining the outcome of collective decision-making processes. Among these causal factors, a significant share of research works has focused on the behaviours that each individual follows to develop and update her opinion. It has been shown that decision-makers with limited perceptual and cognitive resources can follow simple rules to make, as a group, complex collective decisions~\cite{robinson2011simple,sasaki2012groups,reina2018psychophysical}. Among these simple rules, there are those based on social feedback, like the voter model~\citep{masuda2010heterogeneous,moretti2013generalized} where each individual simply copies the opinion of a randomly selected individual within her social network, and the majority model~\cite{galam1986majority,lambiotte2008majority} where each individual selects the option held by the majority of the individuals in her social network. Let us observe that the error in processing social information in the former case of sampling a single randomly chosen opinion is larger than in the latter case of aggregating all available opinions.

This study aims to unravel the mutual influence of psychological, social, and environmental factors on  the outcome of a collective decision-making process. It focuses specifically on the best-of-$n$ scenario with $n=2$ options and an asymmetric condition, i.e., the two options have different costs and/or different benefits~\citep{de2020zealots,talamali2021less,scheidler2011dynamics,prasetyo2020effect}. The best-of-$n$ problem requires selecting which option, out of $n$ available ones, is the best alternative \cite{valentini2017best}. 
Considerable work has been dedicated to the study of the best-of-$n$ problem, in order to unveil opinion dynamics characterising a wide range of phenomena, from political polarisation~\citep{bottcher2020competing}, to the spread of rumours~\citep{difonzo2013rumor} misinformation~\citep{masi2021robot}. We focus on an asymmetric best-of-$2$ scenario since we intend to investigate which factors among those that we modelled hinder the group from a consensus on the best option. In particular, we provide an analysis of the combined effects of the following parameters (which we define in detail in the next paragraph): i) the pooling error, ii) the number of zealots present in the population, iii) the option qualities, iv) the social network structure, and v) the degree centrality of zealot agents.

The pooling error reflects the mistakes introduced by individuals when aggregating the opinions of others. In the given context, these errors may be influenced by the amount of cognitive effort individuals invest in the pooling process. In our model, the pooling error is correlated to the number of group members that each agent samples to develop her own opinion~\citep{CarlettiEtAl2023}. One can recast such error into a cognitive load, namely the lower the cognitive load the larger the pooling error. For example, an agent can face limited cognitive resources and update her opinion based on a peer randomly chosen within her social network (i.e., the voter model). On the other hand, the cognitive load increases when the agent samples a larger number of peers within her social network (e.g., the majority model), and correspondingly the pooling error will decrease. In our model, we consider a heterogeneous population, composed of individuals that update their opinions using different rules. While in a homogeneous population, all individuals use the same rule to select and update their opinions, in a heterogeneous population there are sub-groups of individuals that follow different rules. In particular, we consider a heterogeneous population composed of two groups of individuals, susceptible and zealots. Susceptible individuals at each time step update their opinion based on the opinions of all (or a subset) of their peers in their social network. Zealots are asocial individuals that never change their opinion~\citep{Mobilia2003,khalil2018zealots,masi2021robot,Reina:CommPhy:2023}. The option quality refers to combination the cost and benefit that an agent receives from selecting an option. Our scenario is asymmetric because the two options have different qualities. The social network structure refers to the network that describes how decision-makers are connected and how information flows between them~\citep{momennejad2022collective}. Finally, the degree centrality of an agent is the number of her neighbours, which are other agents in her social network. Degree centrality has been defined in network science as one of the measures to quantify the importance of a node within a network and it has often used to understand the role and influence of decision-makers in social networks~\citep{kameda1997centrality,moeinifar2021zealots}. Nodes with a higher degree centrality are considered more influential because they have a greater number of direct connections, enabling them to spread information or exert their influence more efficiently within the network.

The parameters mentioned above have been separately studied in previous research works. For example,~\citet{moeinifar2021zealots} investigates the effect of the position and the number of zealots on the consensus formation and time to reach consensus in both random and scale-free networks in a symmetric opinions scenario. 
The results show that the degree centrality of the zealots has a great impact on the time required for the group to achieve a consensus. Other studies have shown that, in different quality options scenarios, the asymmetry facilitates the formation of consensus towards the option with the higher quality (or lower cost) \cite{Parker2004, reina2017model,Leonard2024}. 
In this paper, we conduct an analysis of a model that we recently proposed \cite{CarlettiEtAl2023} which allows generalising popular previous models of collective decision-making, from models with high cognitive requirements (such as the majority model) to models requiring low cognitive load (e.g., the voter model). There are several works that investigated the dynamics of the voter model \cite{suchecki2005voter,sood2005voter,Redner2019,schneider2009generalized} or the majority model \cite{lambiotte2008majority,galam1986majority,goles2022majority,nguyen2020dynamics,scheidler2011dynamics} on networks. The model we presented in \cite{CarlettiEtAl2023} allows to interpolate between the two models by changing the (continuous) value of the pooling error, showing the existence of a speed-accuracy trade-off. In this work, we extend the model of \cite{CarlettiEtAl2023} and its analysis by considering heterogeneous populations that comprise susceptible and zealot agents. From the methodological side, we make use of mathematical equations (hereafter, the analytical model) to provide general insights and predictions on the decision process outcomes, and of numerical simulations of an agent-based model (ABM), to corroborate the predictions of the analytical model. By looking at populations with different numbers of zealots holding the lowest quality option, we show interesting correlations between the cognitive load, the number of zealots in the population and their degree centrality, in differently connected populations. In particular, we identify the conditions under which zealots favouring the lowest-quality option manage to counterbalance the quality difference and drive the population toward a consensus on the lowest-quality option. We also demonstrate the effect of network heterogeneity, in particular its sparsity, on the conditions for achieving consensus towards the low-quality option.

The rest of this work is organised as follows. In Section~\ref{sec:model}, we describe the agent-based model with social interactions occurring on a scale-free network and in Section~\ref{sec:ResultsABM} we present the results of the numerical simulation of this agent-based model. Section~\ref{sec:mathmod} presents the mathematical model defined by an ordinary differential equation (ODE) allowing us to study the evolution of group opinion taking into account multiple parameters such as the pooling error (cognitive load), the heterogeneity of the population, the option qualities, the connectivity structure, and the degree centrality of the zealot agents. In Section~\ref{Results}, we show the numerical results of this analytical model showing the combined effects of the previous parameters. Finally, we present our conclusions in Section~\ref{cc}.

\section{The Agent-Based Model}
\label{sec:model}
The aim of this section is to introduce the agent-based model, rooted on the parameters previously introduced and to discuss their impact on the system outcome.
We consider the $N$ agents to be the nodes of a scale-free network composed of $L$ undirected edges, each one representing a possible social interaction between two agents and thus a channel to acquire/share information. {We assume  the number of neighbours of a generic node $i$ (i.e., its degree) follows a power-law distribution~\cite{barabasi1999emergence,PSV2004}, $p_k \sim 1/k^{\gamma}$, where $\gamma > 2$.  Note that the closer $\gamma$ to $2$ the more heterogeneous the degree distribution. Nodes with a very large degree can be observed since $\langle k^2\rangle$ becomes unbounded as $\gamma\rightarrow 2$. When $\gamma \gg 3$ the network becomes sparse, very high degree nodes are very rare and the degree spread is well described by finite variance of the degree distribution}. The networks we consider are simple (i.e., at most one edge can connect any two nodes) and connected (i.e., starting from any node it is possible to reach any other node by traversing the network using the available links).

We classify nodes into susceptible (i.e., representing agents capable of changing their opinions) and zealots (i.e., representing agents that never change their opinion). We model a best-of-$n$ problem with $n=2$ options where each node can hold either opinion $A$ or opinion $B$. We also associate to each option a quality, $Q_A>0$ for opinion $A$, and $Q_B>0$ for opinion $B$. The quality defines the strength or the probability with which the option is communicated to the neighbours. Without lack of generality, in the rest of the work, we assume $Q_A=1$, $Q_B\leq Q_A$ and hence the quality ratio $Q := Q_B/Q_A \leq 1$. In this work, zealots hold the opinion with the lowest quality option (i.e., opinion $B$), since we are interested in investigating under which circumstances they can hinder the population from reaching a consensus (i.e., a general agreement) on the option with the best quality.

\begin{figure}[ht!]
\centering
\includegraphics[width=0.5 \textwidth]{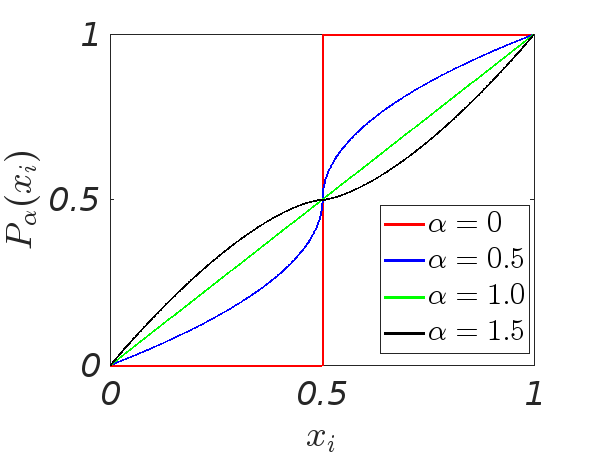}
\caption{Probability function $P_\alpha(x_i)$ for the agent $i$ to change opinion according to the weighted fraction $x_i$ of neighbours holding opinion $A$ for several values of the pooling error parameter, $\alpha$. For $\alpha=0$, our model aligns with the majority model model and for $\alpha=1$, our model aligns with the voter model. In the range $0<\alpha<1$, our model interpolates between the two models, while for $\alpha>1$, agents make large pooling errors and change their opinion with little regard for others' opinions.} 
\label{fig::PF}
\end{figure}

The system evolves asynchronously: at each time step one susceptible individual $i$ is randomly selected with uniform probability from the population. This agent has a probability $P_{\alpha}(x_i)$ as defined in Eq.~\eqref{Px}, of changing her opinion.
\begin{equation}
        P_{\alpha}(x_i) = 
            \begin{dcases}
             \frac{1}{2}-\frac{1}{2}\left(1-2x_i\right)^{\alpha} & \text{if $ 0\leq x_i \leq \frac{1}{2}$}\\
              \frac{1}{2}+\frac{1}{2}\left(2x_i-1\right)^{\alpha}  & \text{if $ \frac{1}{2} < x_i \leq 1$}
            \end{dcases}\, ,
        \label{Px}    
\end{equation}
where, $x_i$ is the weighted fraction of agents connected to agent $i$ holding opinion $A$ or $B$, where the weight is given by the quality ratio $Q$. The parameter $\alpha \geq 0$ is the pooling error (and thus is negatively correlated with the cognitive load); for increasing $\alpha$, agents make larger errors because they use less resources to sample the opinions of their neighbouring agents. More specifically, when $\alpha=0$, from Eq.~\eqref{Px} it follows that agents do not make any sampling error because they change their opinions based on the weighted average of all their neighbours, corresponding thus to the majority model. This action requires a higher cognitive load compared to the case where $\alpha > 0$. For instance, for $\alpha = 1$, agents change their opinions by copying the opinion of a randomly selected neighbour, replicating thus a (weighted) voter model. Therefore, by varying $\alpha$, $P_{\alpha}$ is capable of generalising the two opinion selection mechanisms existing in the literature, i.e., the voter model (see Fig.~\ref{fig::PF}, case $\alpha = 1.0$) and the majority model (see Fig.~\ref{fig::PF}, case $\alpha = 0.0$). The function of Eq.~\eqref{Px} interpolates the cognitive load level in the form of pooling error between the two models in a continuous way (e.g., $\alpha = 0.5$). Additionally, it can also model the case of high pooling errors where the changes are made approximately (e.g., Fig.~\ref{fig::PF}, case $\alpha=1.5$). Using Eq.~\eqref{Px}, we investigate the effect of the pooling error (cognitive load) on the best-of-$2$ problem.

For the randomly selected susceptible agent $i$, the quality-weighted proportion of neighbours holding opinion $A$ (i.e., $n_{i,A}^{\#}$), and the quality-weighted proportion of neighbours holding opinion $B$ (i.e., $n_{i,B}^{\#}$) are computed as follows:
\begin{equation}
        n_{i,A}^{\#} = \frac{Q_A n_{i,A}}{Q_An_{i,A} + Q_Bn_{i,B}} \text{ and }n_{i,B}^{\#} = \frac{Q_B n_{i,B}}{Q_An_{i,A} + Q_Bn_{i,B}}\, ,
\label{nnq}
\end{equation}
where $n_{i,A}$ and $n_{i,B}$ are the number of neighbours of agent $i$ holding opinion $A$ and $B$, respectively. $n_{i,A}^{\#}+n_{i,B}^{\#}=1$, $\forall i$. Let us denote by $k_i=n_{i,A}+n_{i,B}$ the degree of agent $i$. We can rewrite Eq.~\eqref{nnq} as follows:
\begin{equation}
        n_{i,A}^{\#} =   \frac{n_{i,A}/k_i}{(1 - Q)n_{i,A}/k_i + Q} \quad \text{and} \quad
        n_{i,B}^{\#} = 1 - n_{i,A}^{\#}\, .
    \label{nnq1}
\end{equation}
To distinguish between susceptible voters with opinion $B$ and zealots, let us introduce $z_{k}$ to represent the number of zealots with opinion $B$ and with a degree $k$. Therefore, the total number of neighbours of agent $i$ holding opinion $B$ can be written as $n_{i,B} = s_{i,B} + z_{i,B}$ and then, $k_i=n_{i,A}+s_{i,B} + z_{i,B}$.

In the process of changing opinions, if the selected agent $i$ is susceptible and has opinion $A$, then with probability $P_\alpha(n_{i,B}^\#)$ the agent changes her mind and commits to opinion $B$, and with probability $1 - P_\alpha(n_{i,B}^\#)$ she remains committed to opinion $A$. If instead the agent has opinion $B$, then with probability $P_\alpha(n_{i,A}^\#)$ the agent changes her mind and commits to opinion $A$, and with probability $1 - P_\alpha(n_{i,A}^\#)$ she remains committed to opinion $B$. Let us observe that because of the functional form of Eq.~\eqref{Px} and because $n_{i,A}^{\#}+n_{i,B}^{\#}=1$, we can conclude that $P_\alpha(n_{i,A}^{\#})+P_\alpha(n_{i,B}^{\#})=1$.

\section{Numerical results of the agent-based-model} 
\label{sec:ResultsABM}
The objective of this section is to present the results of the numerical simulation of the agent-based model with social interactions occurring on a scale-free network. In this section, we limit our analysis to studying the combined effects of three factors: i) the number of zealots ($z$), ii) their location in the network, and iii) the  pooling error ($\alpha$), while we keep constant the network structure and the option's quality ratio $Q=Q_B/Q_A = 0.9$.

We consider a system composed of $N=1000$ agents whose social interactions are described by a scale-free network obtained by using the Barab\'asi-Albert algorithm~\cite{barabasi1999emergence} with parameter $m=8$ (i.e., a network in which the minimum degree is $k_{min}=8$). The exponent of the power law of such a network is $\gamma=3$. In our experimental plan, we have $20$ different parameters conditions obtained by considering $4$ different values of $\alpha \in \{0.0, 0.5, 1.0, 1.5\}$, and $5$ different quantities of zealots, $z\in \{0, 32, 64, 128, 150\}$.
We consider cases where the zealots, committed to opinion $B$, have similar locations in the social network. That is, they can be peripheral agents (i.e., associated with nodes with the smallest degree $k_{\text{min}} = 8$), or agents associated to nodes with an intermediate degree, or a large degree. The opinion qualities have been fixed to $Q_A=1$ and $Q_B=0.9$.
The initial opinions are randomly distributed with half of the agents holding opinion $A$ ($n_A(0)=N/2$) and half opinion $B$ ($n_B(0)=N/2$), and the maximal simulation time has been set to $T_{max}=400\, 000$. Each experimental plan is repeated $30$ times by changing each time the seed of the pseudo-random number generator.

\begin{figure}[htp!]
        \centering
        \includegraphics[width=0.5\textwidth]{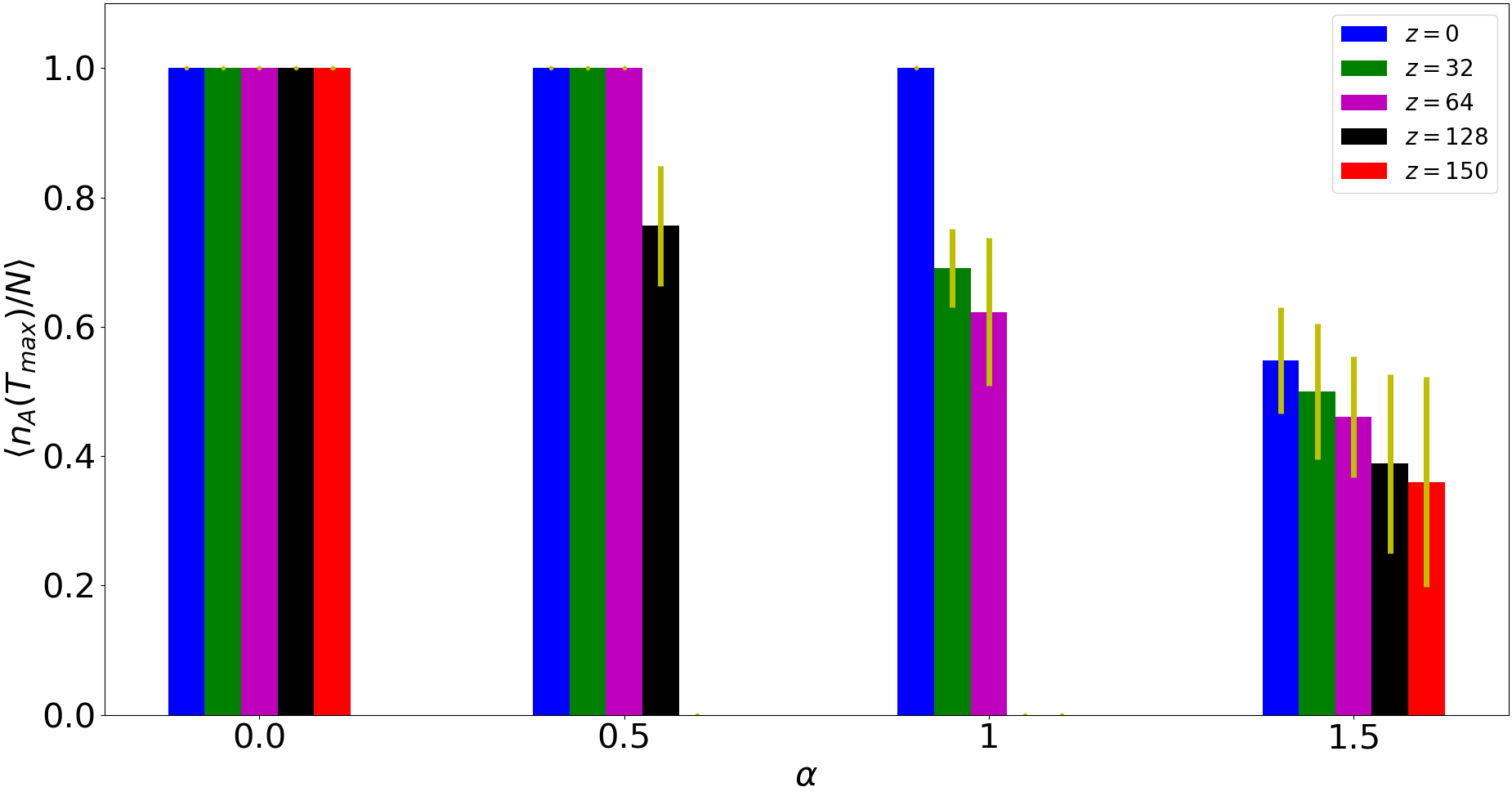}
        \caption{We report the average final fraction of agents with opinion $A$ in the population of $N=1000$, interacting on a Barab\'asi-Albert network with $m=8$, and the standard deviation (yellow line), resulting from  $30$ independent simulation runs. Different values of the pooling error $\alpha$ are considered (on the x-axis) together with a varying number of zealots (different colours). zealots are committed to opinion $B$ are set into leaves nodes, i.e., $k_i=8$. $Q_A=1$ and $Q_B=0.9$ for options $A$ and $B$, respectively.}
        \label{fig::FF1H}
\end{figure}
In Fig.~\ref{fig::FF1H}, we report the results of the numerical simulations performed by adopting the strategy of introducing zealots with low degree centrality. More precisely we illustrate the average final fraction of susceptible agents holding opinion $A$ computed over $30$ independent replicas together with the standard deviation (vertical yellow lines). In the absence of zealots (blue bars, $z=0$, in Fig.~\ref{fig::FF1H}), the system always converges to a consensus toward opinion $A$ for $\alpha \in \{0.0, 0.5, 1.0\}$. Indeed the standard deviations associated to those values of $\alpha$ are zero, thus, the yellow lines reduce to points. On the other hand, for $\alpha=1.5$, and still $z=0$, the group faces a decision deadlock where the population splits into two coexisting groups of agents, one having opinion $A$ and the other opinion $B$, the average fraction of agents committed to $A$ is $0.5$ and the standard deviation is relatively small. In other words, due to high pooling error ($\alpha>1$), agents fail to coordinate with each other and the population remains polarised and unable to reach a consensus for any alternative, especially when the quality difference between the two options is small. With the introduction of zealots in the population, for $\alpha \in \{0.5, 1.0, 1.5\}$, we notice that the increase in the number of zealots lowers the final proportion of susceptible agents with opinion $A$. In the cases $\alpha \in \{0.5, 1.0\}$, a sufficiently large number of zealots ($z=150$) is even able to drive consistently the system toward a full consensus to opinion $B$ (observe that in Fig.~\ref{fig::FF1H} the red bars for $\alpha= 0.5$ and $\alpha=1.0$ vanish, and the standard deviation is reduced to zero). For $\alpha=1.5$, agents with opinion $A$ persist in the population despite the presence of (many) zealots, this is due to the large pooling error that keeps the population polarised and deadlocked at indecision, unable to choose one opinion over the other. This last observation holds true even for a large number of zealots (see the cases $z=128$ and $z=150$), the average fraction of $A$ agents decreases and the standard deviation increases, indicating a larger variability among the different runs. For $\alpha=0.0$, indicating the absence of pooling error (majority model), the existence of zealots does not change the outcomes observed without the presence of zealots (see all the non-blue bars in Fig.~\ref{fig::FF1H}), again we can observe a null standard deviation.

To study the combined impact of the number of zealots and their location in the social network, we have performed a further series of simulations in order to verify whether the position of the zealots within a scale-free network of $N=1000$ nodes, generates a similar effect to the one observed with respect to their number. We considered $60$ different experimental conditions, given by $5$ different amount of zealots, $z \in \{0, 1, 4, 8, 16\}$, $3$ values of $\alpha\in \{0.5, 1.0, 1.5 \}$, and $4$ different strategies to locate zealots in the network according to agent's degree $k$. With the first strategy, zealots correspond to nodes with degree $k_{min}=8$.  
With the second strategy, zealots correspond to nodes with a degree $15 \leq k \leq 25$. With the third strategy, zealots correspond to nodes with degree $35 \leq k \leq 45$. Finally, with the fourth strategy, zealots correspond to nodes with a degree $k \geq 60$. Note that, in the last three strategies, we first associate zealots with nodes with highest degree and we progressively select nodes with lower degree only once there are no more available nodes with the larger degree in the network. As in the previous set of simulations, the quality of option $A$ is set to $Q_A=1$, and that of opinion $B$ to $Q_B=0.9$; the initial opinions are randomly distributed with half of the agents holding opinion $A$ ($n_A(0)=N/2$) and half opinion $B$ ($n_B(0)=N/2$). We repeated each experiment $30$ times by using different random seeds, and each simulation lasts for $T_{max}=400\, 000$ time steps. 

\begin{figure}[htp!]
\centering
\begin{tabular}{c}
\includegraphics[width=0.475\textwidth]{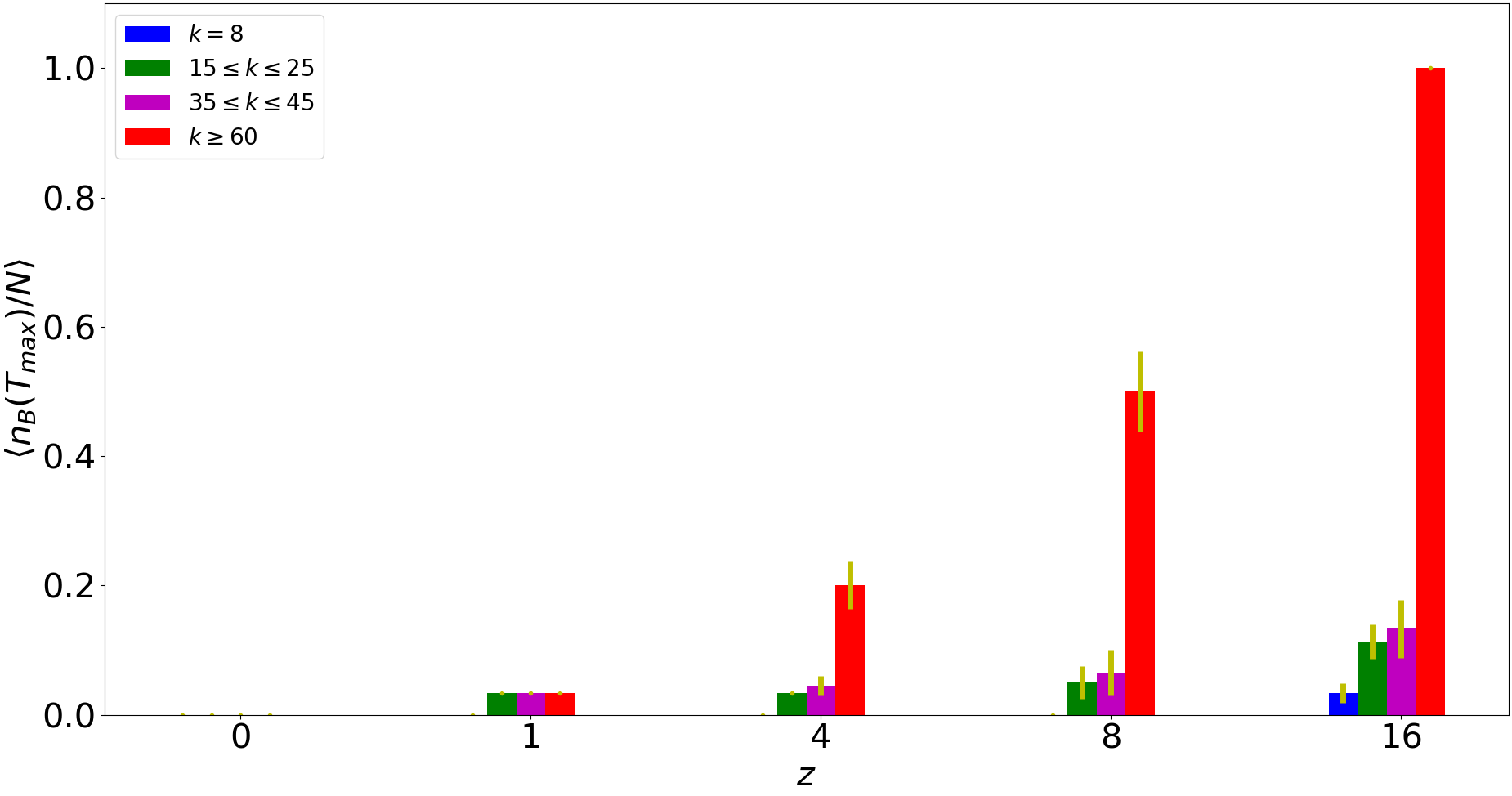}\\
(a)\\
\includegraphics[width=0.475\textwidth]{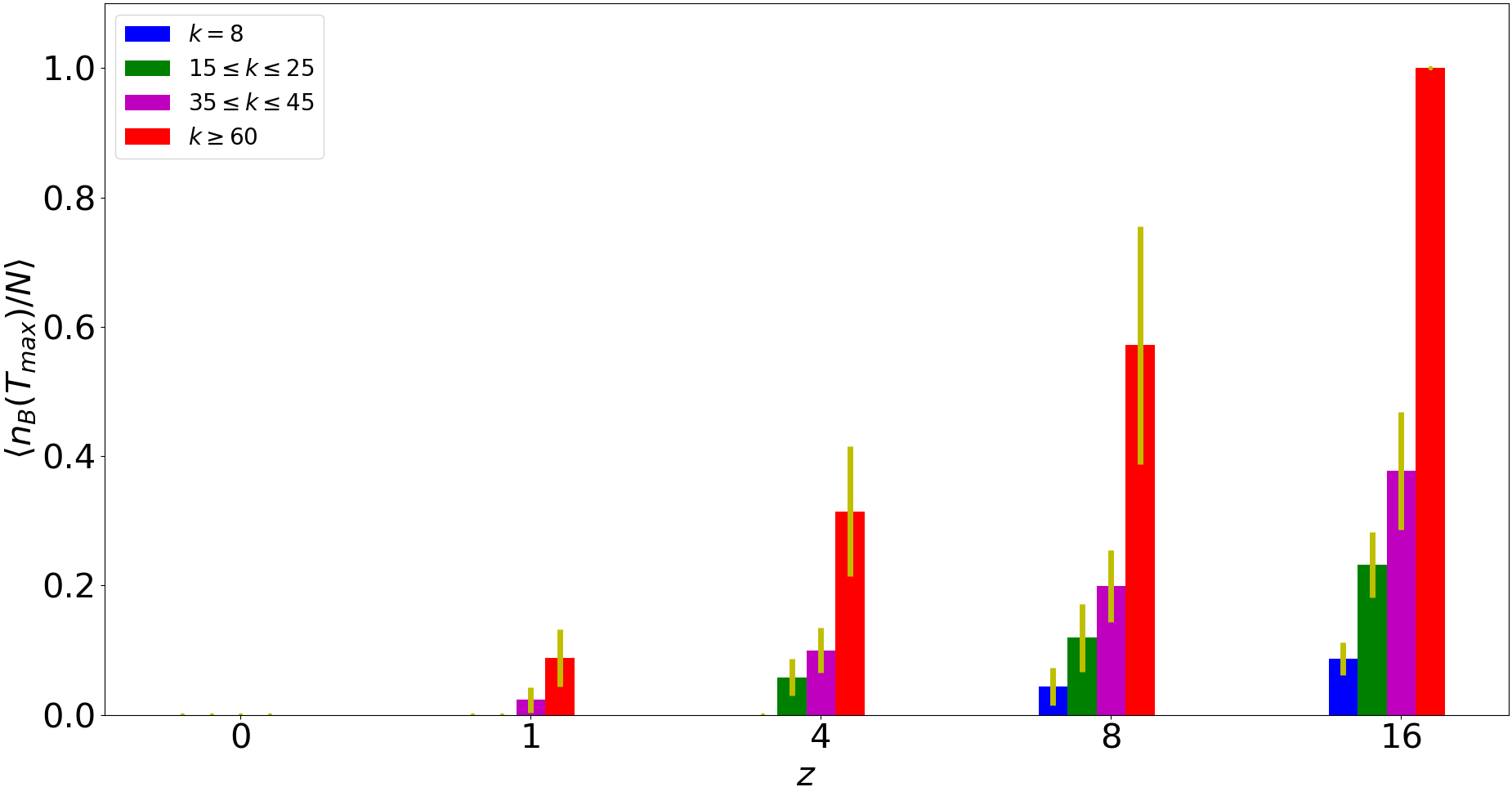}\\
(b)\\
\includegraphics[width=0.475\textwidth]{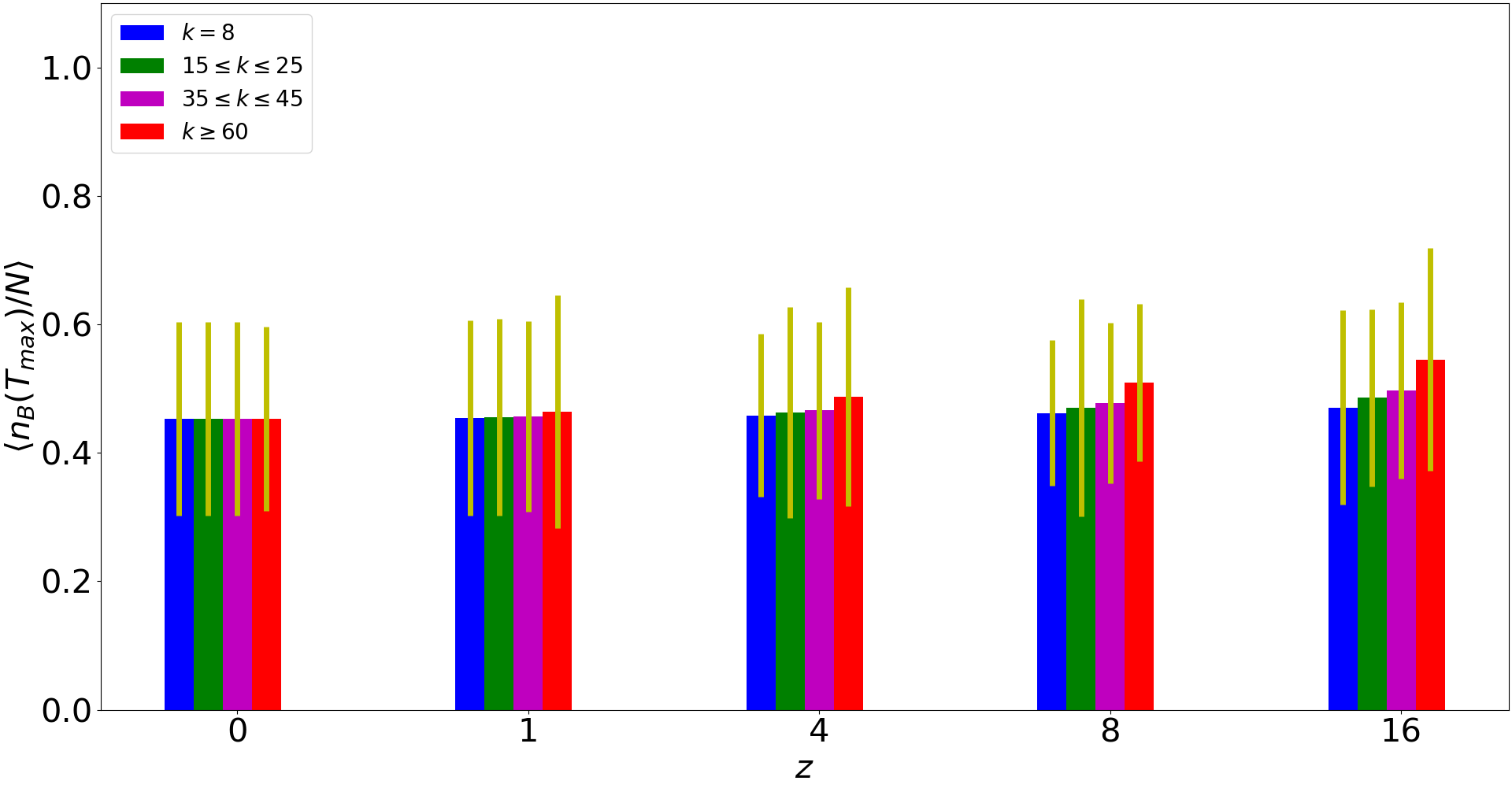}\\
(c)
\end{tabular}
\caption{We report the average final fraction of agents with opinion $B$ in the population of $N=1000$, interacting on a Barab\'asi-Albert network with $m=8$, and the standard deviation (yellow line), resulting from  $30$ independent simulation runs. Panel (a) corresponds to pooling error $\alpha=0.5$, panel (b) to $\alpha=1.0$, and panel (c) to $\alpha=1.5$. The colours of the bars refer to zealots with different degrees of centrality (see text). $Q_A=1$ and $Q_B=0.9$ for options $A$ and $B$, respectively.}
\label{fig::FF1C}
\end{figure}

The results of these simulations are shown in Fig.~\ref{fig::FF1C}, where we report the average fraction of agents holding opinion $B$ at the end of the simulations, and the associated standard deviations (vertical lines) for each considered value of $z$. For $\alpha=0.5$ (see Fig.~\ref{fig::FF1C}a) and for $\alpha=1.0$ (see Fig.~\ref{fig::FF1C}b), we notice two clear trends: first, with the exception of $z=1$, for any given number of zealots within the population, the higher their degree the higher the final proportion of susceptible agents terminating with opinion $B$. Second, for zealots with a given degree centrality, the higher the number of zealots, the higher the final proportion of susceptible agents terminating with opinion $B$. The increase in the proportion of susceptible agents committed to opinion $B$ at the end of the simulations appears to be non-linearly correlated with the number of zealots. These trends indicate that, in an asymmetric best-of-$2$ problem, given a fixed number of zealots committed to the lowest quality option, their capability to influence the opinion dynamics of susceptible agents using opinion selection strategies with either small or large pooling error, is determined by both their quantity within the population and their degree centrality. Note that, for the values of the considered parameters, when zealots correspond to small degree nodes, approximately $z=150$ zealots are required to converge to a consensus on the lowest quality opinion $B$, while for the same values, at least $z=128$ zealots are needed in the case of $\alpha=1$ (see Fig.~\ref{fig::FF1H}). In the same vein, in the case of $\alpha=1.5$ (see Fig.~\ref{fig::FF1C}c) for any number of zealots, and regardless of their degree, the population always splits into two groups of (almost) equal sizes with different opinions. As previously stated, the standard deviation increases with the number of zealots and with the pooling error.

The results shown in this section indicate that adding zealots with the lower quality option can drive the system to a consensus toward the latter. More importantly, this outcome is generated by a smaller number of zealots if they have a higher degree. The analysis has been performed on a scale-free network obtained by using the Barab\'asi-Albert algorithm. The goal of the next section is to move a step forward in confirming such conclusions in the whole generality in the framework of scale-free networks by using the Heterogeneous Mean-Field theory.

\section{Mathematical model}
\label{sec:mathmod}
We build a mathematical model defined by an ordinary differential equation (ODE) to look at the combined effects of the pooling error, the ratio of the opinion qualities $Q:=Q_B/Q_A$, the fraction of zealots, and the network structure $\gamma$, on the evolution of opinion dynamics in the best-of-$2$ problem.

To make some analytical progress we rely on the Heterogeneous Mean Field assumption~\cite{pastor2015epidemic,PSV2001,CPSV2007} (HMF), namely we hypothesise that nodes with the same degree are dynamically equivalent. Therefore, nodes are grouped into degree classes, more precisely we define $A_k$ (resp. $B_k$), as the number of nodes with degree $k$ and opinion $A$ (resp. opinion $B$). To distinguish between susceptible agents with opinion $B$ and zealots, we introduce $Z_{k}$ to denote the number of zealots with opinion $B$ and degree $k$. Therefore, by letting $N_k$ to denote the total number of nodes with degree $k$, we have:
\begin{equation}
	    A_k + Z_{k} + S_{k} = N_k \, , 
	\label{H1}
\end{equation}
where $S_{k}$ denotes the number of susceptible agents having opinion $B$ and degree $k$.  $a_k = A_k / N_k$ is the fraction of agents having opinion $A$ and degree $k$; $b_k = S_{k} / N_k$ is the fraction of susceptible agents having opinion $B$ with degree $k$, and $\zeta_k = Z_{k} / N_k$ the fraction of zealots with degree $k$. Therefore, for all $k$,
\begin{equation}
a_k + b_{k} + \zeta_{k} = 1\, . 
\label{Ea2}
\end{equation}

The goal of the HMF is to derive an ODE ruling the evolution of $a_k$ and $b_k$. In particular, we extend the model recently developed in~\cite{CarlettiEtAl2023} by adding zealots; in this way, we obtain the analytical model described by 
\begin{eqnarray}
\label{Ea3}
\frac{d\langle a \rangle}{dt} &=& -\langle a \rangle+\sum_k q_k (1-\zeta_{k+1})\sum_{\ell=0}^{k+1} \binom{k+1}{\ell} \langle a\rangle^{k+1-\ell} \times \notag\\
& & \left(1-\langle a\rangle\right)^\ell P_\alpha\left( \frac{k+1-\ell}{k+1-\ell +\ell Q}\right)\, ,
\end{eqnarray}
where we define $\langle a \rangle = \sum_k q_k a_{k+1}$, being $q_k$ the probability for a node to have an excess degree $k$, namely
\begin{equation*}
    q_k = \frac{\left(k+1\right)p_{k+1}}{\langle k \rangle} \quad \forall k\geq 0\, ,
\end{equation*}
with $\langle k \rangle = \sum_k kp_k$ the average node degree and $p_k$ the probability a generic node has degree $k$. Eq.~\eqref{Ea3} contains the relevant parameters of the model, the zealots ($\zeta_k$), the model of opinion dynamics ($P_\alpha$) incorporating the pooling error $\alpha$, the opinion quality ratio ($Q$), and the network structure ($q_k$). The aim of the next subsection is to determine the equilibria of the HMF equation and their stability, and thus the system fate. Let us observe that, similarly to what we did in~\cite{CarlettiEtAl2023}, by knowing $\langle a\rangle(t)$ from Eq.~\eqref{Ea3} we can obtain the evolution of $a_k$ for all $k$ by using the following equation:
\begin{eqnarray*}
          \frac{da_k}{dt} &=& - a_k + \left( 1 - \zeta_{k} \right)   \sum_{\ell=0}^{k-1} \left( \begin{array}{c} k \\ \ell \\ \end{array} \right) \langle a \rangle ^{k-\ell} 
      \left( 1 - \langle a \rangle \right)^\ell \times \notag\\
      & & P_\alpha \left( \frac{k-\ell}{k-\ell+\ell Q} \right) \, .
\end{eqnarray*}

\subsection{Equilibria of the analytical model and their stability} 
\label{method::Equilibria}
The equilibria of the system are obtained by setting the right-hand side of Eq.~\eqref{Ea3} equal to zero. Let us thus define the function $f_{\alpha}(a)$
\begin{eqnarray}
f_{\alpha}(a):&=&-\langle a \rangle +\sum_k q_k (1-\zeta_{k+1})\sum_{\ell=0}^{k+1} \binom{k+1}{\ell} \langle a \rangle^{k+1-\ell} \times \notag\\ & & \left(1- \langle a \rangle\right)^\ell P_\alpha\left( \frac{k+1-\ell}{k+1-\ell +\ell Q}\right)\, ,
    \label{Ea6}
\end{eqnarray}
hence by denoting $\langle a^*\rangle$ a system equilibrium, we have by definition
\begin{equation*}
    f_{\alpha}(\langle a^*\rangle)=0\, .
\end{equation*}
A direct inspection of Eq.~\eqref{Ea6} allows to demonstrate that $f_\alpha(0)=0$, hence $\langle a^*\rangle=0$, i.e., the absence of agents holding opinion $A$ represents an equilibrium of the system. Conversely, $f_\alpha(1)=-\sum_k q_k\zeta_{k+1}\neq 0$, signifying that the presence of zealots (with opinion $B$) prevents the system from converging to a situation where all susceptible agents hold opinion $A$. Finally, the existence of a nontrivial solution $0 < \langle a^* \rangle < 1$ to the equation $f(\langle a^* \rangle) = 0$  indicates an equilibrium for the coexistence of opinions $A$ and $B$ in the network.

The stability of those equilibria can be determined by evaluating the derivative of the function $f_\alpha$ at those points. This analysis will be detailed in the following section, where we will also explore the effects of the key model parameters.

\section{Results of the HMF model}
\label{Results}
In this section, we present the results obtained for the analytical model described in the previous section. As previously noted, our focus lies on the influence of the pooling error $\alpha$, the network structure encapsulated into the exponent $\gamma$ of the power law, and the ratio of zealots present in the population and their degree centrality, more precisely if they sit onto hubs or leaves nodes. To place zealots in hubs, we set $\zeta_{k}=1$ for all $k\geq k_M$, for some sufficiently large $k_M>0$; this accounts to add into the model an average number of zealots equal to $Z_{tot}=\sum_{k\geq k_M} N_k\sim \sum_{k\geq k_M} N c_\gamma /k^\gamma$, where $c_\gamma$ is a normalisation constant such that $\sum_k p_k=1$ and $N$ is the total number of nodes in the network. In the scenario where zealots are assumed to be on leaf nodes, for a fair comparison with the prior condition, we maintain the same number of zealots as positioned in the hubs. This is achieved by assuming $\zeta_{k_{min}}=Z_{tot}/N_{k_{min}}$, where $k_{min}>0$ represents a sufficiently small degree. Specifically: 
\begin{eqnarray*}
 \zeta_{k_{min}}&=&\frac{Z_{tot}}{N_{k_{min}}}\sim\frac{Z_{tot}}{N p_{k_{min}}} \notag\\
 & = & k_{min}^\gamma \sum_{k\geq k_M} \frac{1}{k^\gamma}\sim  \frac{k_M}{\gamma-1}\left(\frac{k_{min}}{k_{M}}\right)^\gamma\, .
\end{eqnarray*}
Let us observe that the above strategy implies that $\zeta_k$ can be interpreted as the probability to find a zealot into the class of nodes with degree $k$; the final number of added zealots will be always finite, indeed in any network realisation, e.g., by using the configuration model, there is a finite number of nodes with degree larger than $k_M$ and thus $Z_{tot}$ is also a finite quantity. 

\begin{figure*}[ht!]
\vspace{-2.0cm}
\centering
\includegraphics[width=1.05\textwidth]{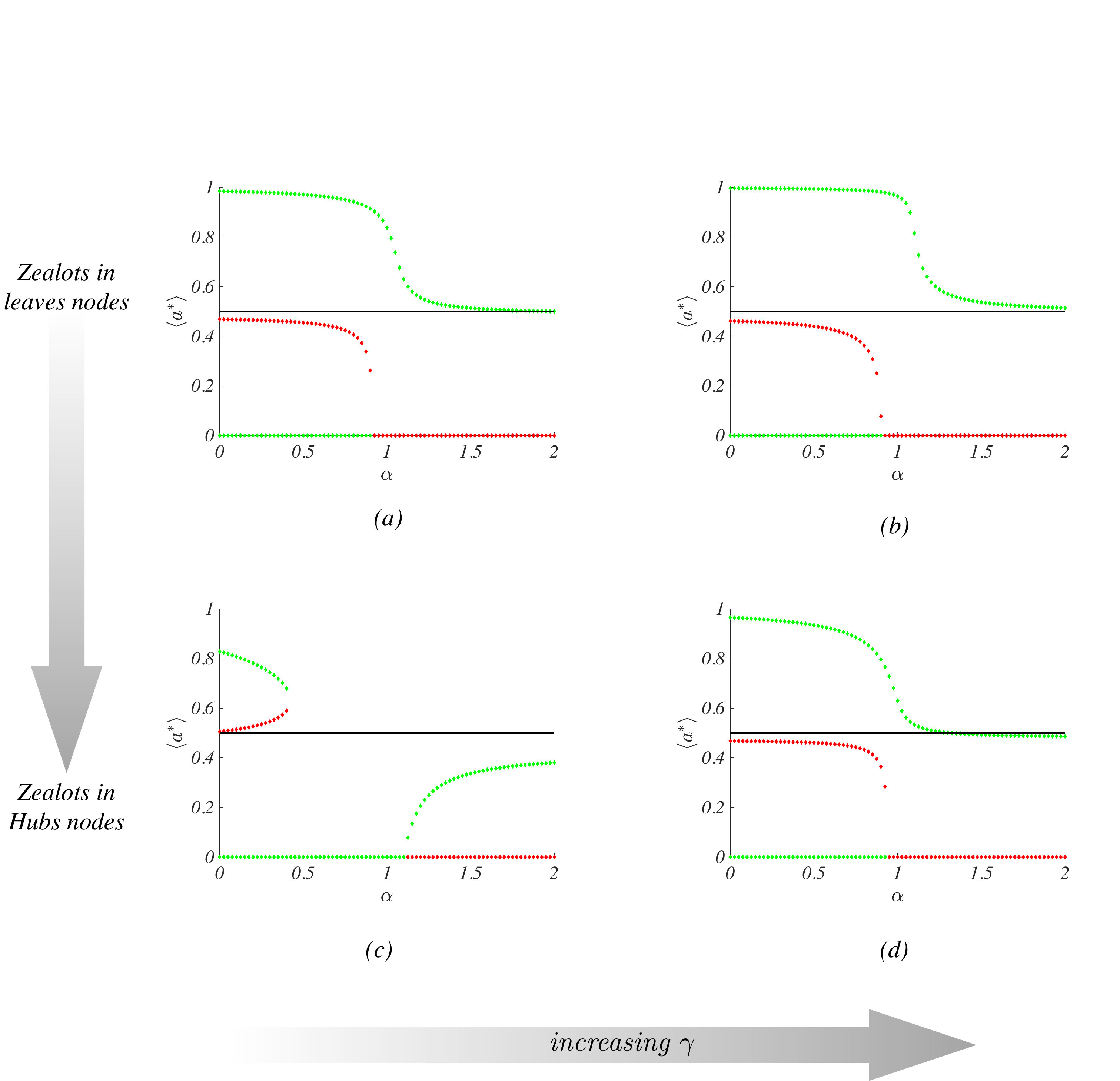}
\caption{Bifurcation diagrams of the HMF. We report the equilibria $\langle a^*\rangle$ of Eq.~\eqref{Ea3} as a function of the pooling error $\alpha$ for $Q=0.9$. Stable equilibria, i.e., associated to $f^\prime_\alpha(\langle a^*\rangle)<0$, are coloured in green while unstable ones, i.e., associated to $f^\prime_\alpha(\langle a^*\rangle)>0$, are coloured in red. Top panels (a) and (b), correspond to zealots set into leaves nodes, i.e., $k_i=8$, while bottom panels (c) and (d) correspond to zealots set into the large degree nodes, more precisely into nodes with $k_i\geq 60$. The underlying support is a scale--free network, with $\gamma=2.5$ (panels (a) and (c)) and $\gamma=3.5$ (panels (b) and (d)).}
\label{fig::Res}
\end{figure*}

Fig.~\ref{fig::Res} summarises our main results. The quality of options $A$ and $B$ are set as in the previous section, i.e., $Q_A=1$ and $Q_B=0.9$, resulting in $Q=Q_B/Q_A=0.9$. We then proceed to vary the power law exponent $\gamma$ and the location of zealots in the network. Subsequently, we (numerically) determine the zeros of the function $f_\alpha$ for varying values of $\alpha$ within the range of $[0,2]$, enabling us to derive the system equilibria. Once the latter have been found, we evaluate the derivative of $f_\alpha$ and we determine its sign, if $f_\alpha(\langle a^*\rangle)>0$ then the equilibrium $\langle a^*\rangle$ is unstable and marked with red points in Fig.~\ref{fig::Res}. 
On the other hand, if $f_\alpha(\langle a^*\rangle)<0$ then the equilibrium $\langle a^*\rangle$ is stable and we represent it in green. The two top panels (see Fig.~\ref{fig::Res}(a) and (b)) refer to the strategy consisting of setting the zealots in the leaves (here $k_{min}=8$), and the two bottom panels (see Fig.~\ref{fig::Res}(c) and (d)) refer to the opposite strategy with the zealots in the hubs, $k_M=60$; this strategy is thus the same of the one used for the ABM in Section~\ref{sec:ResultsABM}. Moving from left to right we increase $\gamma$, passing from $\gamma=2.5$ (Fig.~\ref{fig::Res}(a) and (c)), to $\gamma=3.5$ (Fig.~\ref{fig::Res}(b) and (d)).

Several conclusions can be drawn from those results. For large enough values of $\alpha$, the system always sets into a state where opinions $A$ and $B$ coexist, leading to a state of decision deadlock where the population is unable to choose one opinion over the other. The larger is $\alpha$ the closer is the stable equilibrium to $0.5$; this behaviour is independent of the strategy used to place the zealots or the network structure determined by the exponent $\gamma$. Hence, a too-large pooling error $\alpha$, and thus a small cognitive load, prevents the agents from reaching a consensus for their alternative regardless of the position of zealots. 

For intermediate values of the pooling error $\alpha$, e.g., close to $\alpha=1$, placing the zealots into the low-degree nodes allows the system to reach a consensus for the best option (see panels 4(a) and 4(b)) irrespective from the value of $\gamma$, indeed there is a stable equilibrium $\langle a^*\rangle$ very close to $1$. This behaviour completely changes once zealots are set into hubs nodes (see panels 4(c) and 4(d)), indeed for small $\gamma$, i.e., a network with a pronounced degree heterogeneity, there exists an interval of $\alpha \in [0.41, 1.10]$ (see panel 4(c)) values for which the unique stable equilibrium is $\langle a^* \rangle=0$, the HMF predicts thus the system to converge to the opinion with lower quality, $B$. By increasing $\gamma$, i.e., dealing with a more homogeneous network, we recover once again convergence to a large fraction of agents committed to opinion $A$.

Finally, for very low values of the pooling error $\alpha$, i.e., $\alpha \sim 0$, placing zealots in low-degree nodes allows the system to reach a consensus for the best option. Therefore, a stable equilibrium $\langle a^* \rangle \sim 1$ is achieved for any value of $\gamma$ (see panels 4(a) and 4(b)). However, when these zealots are placed on hubs in a network with much greater heterogeneity, the stable equilibrium representing a majority of agents with opinion $A$ reduces to values lower than 1. Thus, while most agents adopt the higher-quality opinion, there is also a significant proportion of agents with opinion $B$. This is shown in Fig.~\ref{fig::Res}(c) by the stable equilibrium point $0.5 < \langle a^* \rangle < 1$. Furthermore, when the network structure becomes more homogeneous (e.g., $\gamma=3.5$), the presence of zealots does not prevent the system from reaching a consensus towards the higher-quality option, thereby reaching the stable equilibrium point $\langle a^* \rangle = 1$ (see panel 4(d)).

\begin{figure*}[htp!]
\vspace{-1cm}
\centering
\includegraphics[width=1.05\textwidth]{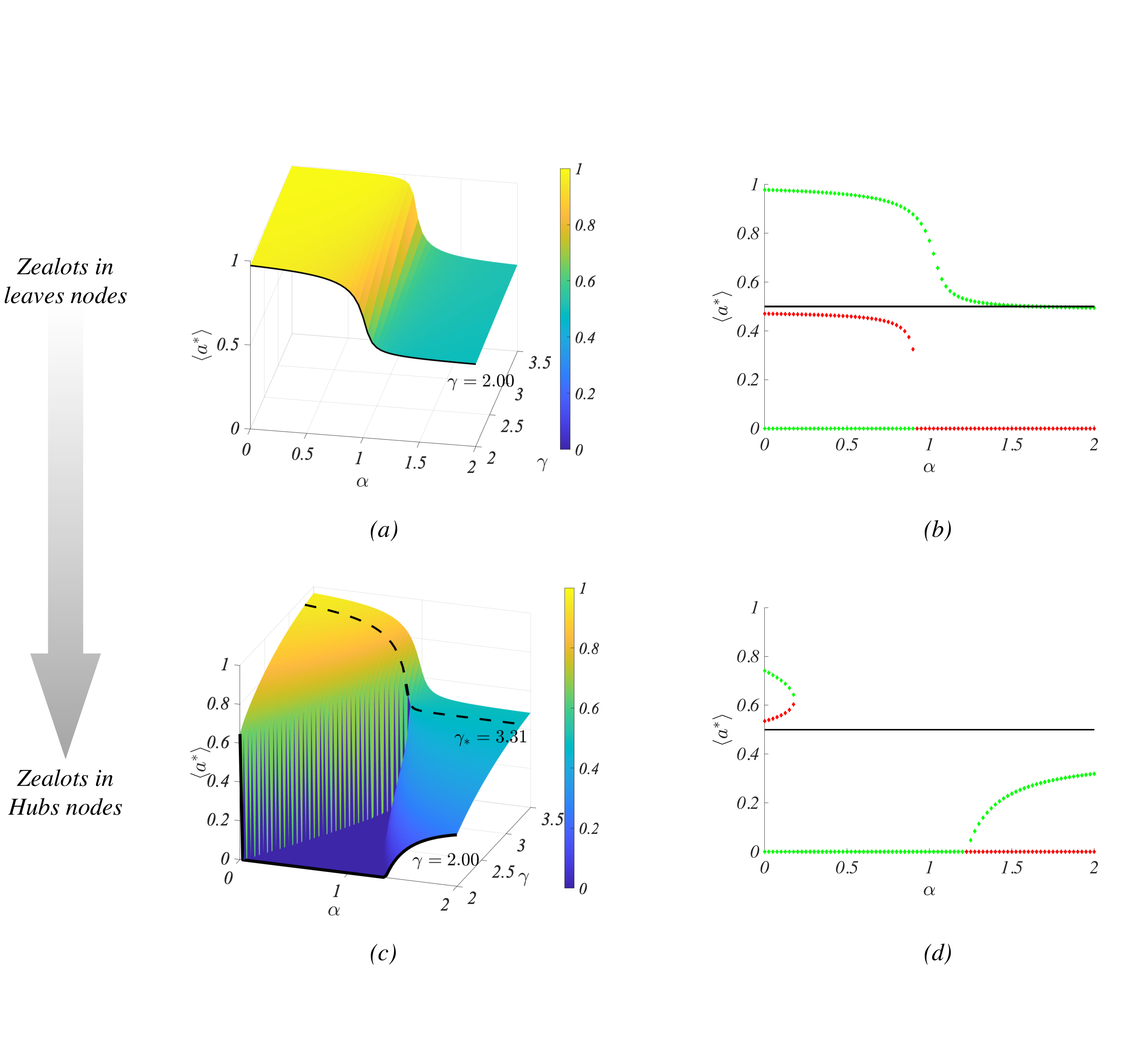}
\vspace{-2cm}
\caption{Bifurcation diagrams of the HMF. We report the equilibrium $\langle a^*\rangle$ given by Eq.~\eqref{Ea3} as a function of $(\alpha,\gamma)$ for a fixed value of $Q=0.9$. Top panels (a) and (b) correspond to the strategy of placing zealots into leaves nodes, i.e., $k_i=8$, while in the bottom panels (c) and (d), zealots are assigned to large degree nodes, more precisely to nodes such that $k_i\geq 60$. In panel (c) we can observe the existence of a critical value $\gamma_*$ of the power-low exponent $\gamma$, below which the equilibrium $\langle a^*\rangle =0$ is the unique stable one, for a given range of $\alpha$. Panels (b) and (d) show the equilibria for a fixed value of $\gamma=2.2<\gamma_*\sim 3.31$.}
\label{fig::Res3D}
\end{figure*}

To achieve a broader understanding of the complex relationship between the parameters, we studied the equilibrium $\langle a^*\rangle$ as a function of the pooling error $\alpha$ and the exponent $\gamma$, while maintaining constant the ratio of the opinion qualities $Q=0.9$ (we report the results in Fig.~\ref{fig::Res3D}). Furthermore, for each examined scenario, we assessed the influence of the strategy involving the placement of zealots on leaf nodes (see Fig.~\ref{fig::Res3D}(a) and (b)) or on hubs nodes (see Fig.~\ref{fig::Res3D}(c) and (d)). In Fig.~\ref{fig::Res3D}(a) and (c) we represented by a colour code (yellow high values of $\langle a^*\rangle$ close to $1$ and blue $\langle a^*\rangle \sim 0$) the equilibrium reached by the system starting from an initial population with half agents holding opinion $A$ and half opinion $B$ (note that half of agents holding opinion $A$ are all susceptible agents while the other half, holding opinion $B$, also includes the zealots). One can observe a striking difference between Fig.~\ref{fig::Res3D}(a) corresponding to zealots placed into leaves nodes, here $k_{min}=8$, with respect to the Fig.~\ref{fig::Res3D}(c), where the zealots have been set into hubs nodes, here $k_M\geq 60$. In the former case, the equilibrium $\langle a^*\rangle$ is almost independent from $\gamma$ and the system exhibits two main behaviours: for $\alpha \lesssim 1$ the whole group converges to a consensus to $A$, while for $\alpha \gtrsim 1$ the population faces a deadlock where agents with opinion $A$ and $B$ coexist. On the other hand, once zealots are placed into hubs nodes a third type of dynamics can manifest (see Fig.~\ref{fig::Res3D}(c)): the population can converge toward a consensus for the opinion with the lower quality. As shown in Fig.~\ref{fig::Res3D}(c)), this happens for $\alpha \sim 1$, $Q=0.9$, and for $\gamma \lesssim \gamma_*=3.31$ (the value of $\gamma^*$ has been computed numerically). To better emphasise this behaviour, we report in Fig.~\ref{fig::Res3D}(b) and (d) the position and the stability of the equilibrium $\langle a^*\rangle$ as a function of the pooling error for the value $\gamma=2.2 < \gamma_*\sim 3.31$. Fig.~\ref{fig::Res3D}(b) corresponds to the case where zealots are set into the leaves nodes and the population converges to a (almost) consensus to $A$ for $\alpha \lesssim 1.0$, while for larger values of $\alpha$ the population faces a decision deadlock. On the other hand, once zealots are set into hubs (Fig.~\ref{fig::Res3D}(d)), there exists an interval of values of $\alpha$ for which the group chooses the opinion with the lower quality. Those results support the claim that a population of agents adopting a voter model strategy for social exchange can be driven to adopt the opinion with the lower quality, by zealots placed into hubs of a sufficiently heterogeneous scale-free network, i.e., $\gamma < \gamma_*$.

\begin{figure*}[ht!]
\vspace{-2cm}
\centering
\includegraphics[width=1.05\textwidth]{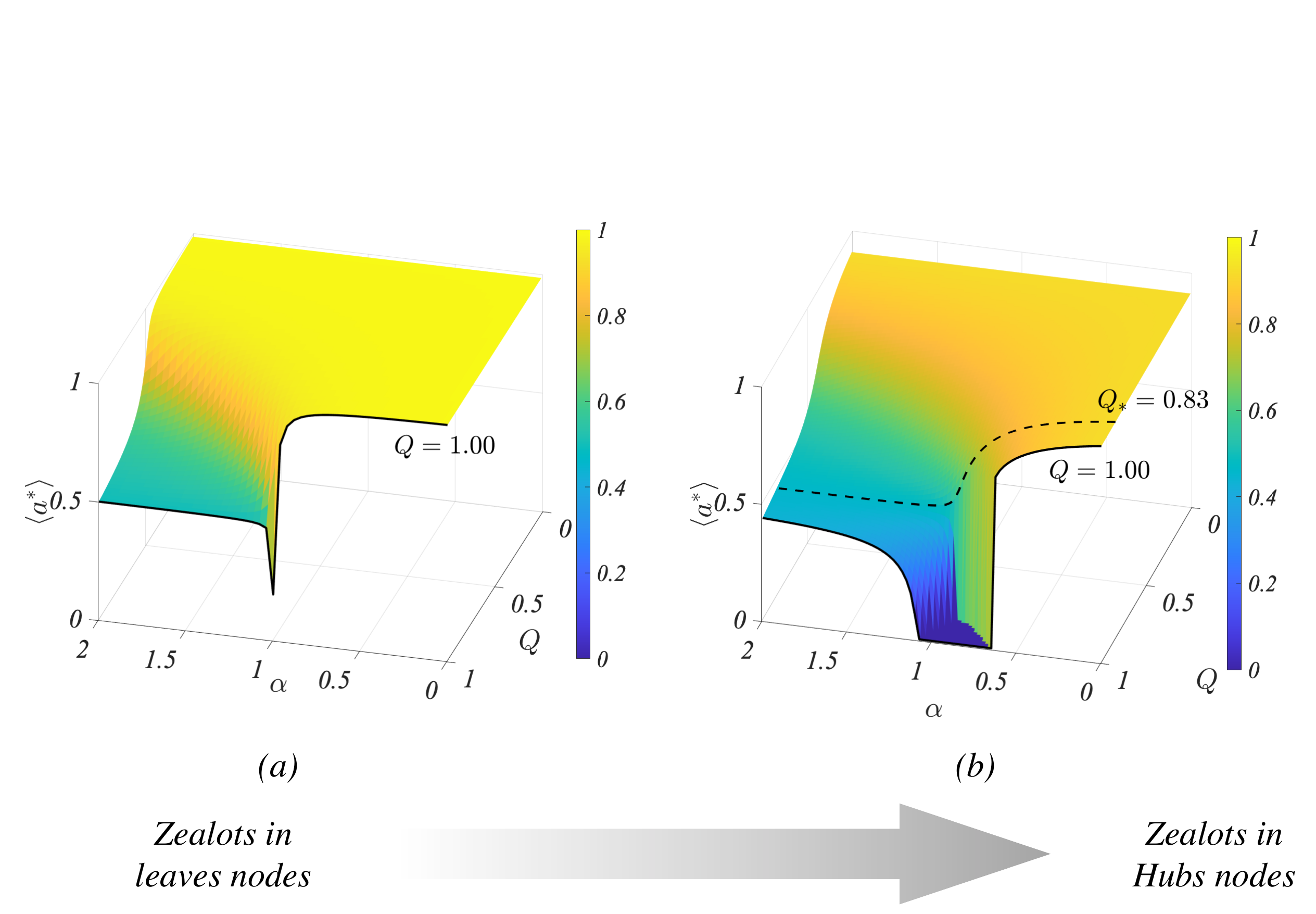}
\caption{Bifurcation diagrams of the HMF. We report the equilibrium $\langle a^*\rangle$ given by Eq.~\eqref{Ea3} as a function of $(\alpha,Q)$ for a fixed value of $\gamma=3.0$. Left panel (a) corresponds to zealots set into leaves nodes, i.e., $k_i=8$, while the panel (b) to the strategy of placing the zealots into the hubs, i.e., nodes with $k_i\geq 60$. Let us observe the presence in panel (b) of a critical value of $Q=Q_*\sim 0.83$, below which the equilibrium $\langle a^*\rangle=0$ is the unique stable one.}
\label{fig::Res3DQ}
\end{figure*}

Fig.~\ref{fig::Res3DQ} shows that similar dynamics can be observed by considering equilibrium $\langle a^*\rangle$ as a function of $(\alpha,Q)$ for a fixed value of $\gamma$. The results presented in Fig.~\ref{fig::Res3DQ}(b) show that in the case of zealots located into hubs, for $Q$ close enough to $1$ (i.e., opinions with very similar qualities) and $\alpha \sim 1$, the population converges to a majority for the option with the lower quality; by decreasing $\alpha$, the system undergoes an abrupt bifurcation passing to a population with a large majority of agents holding an opinion in favour of the best quality. This behaviour cannot be observed if zealots are placed into the leaves nodes (see Fig.~\ref{fig::Res3DQ} (a)). Results presented in panel 6(b) show the existence of a critical value of $Q_*$ (obtained numerically) above which the system can converge toward the lowest quality option for some values of $\alpha$.

Our analytical framework confirmed the numerical results shown in Sec.~\ref{sec:ResultsABM} and generalises the analysis to any scale-free network, under the assumption of heterogeneous mean field.

\section{How much quality is worth a zealot?}
\label{rem:QvsZ}
In the previous sections, we have considered the quality of the options as {\em a priori} measure of the value of one choice over the other, or the strength with which an option is transmitted. In the case of humans the latter role can be played by, e.g., mass media penetration. In the same framework, zealots, can thus be considered as individuals (or bots) repeating constantly one option with the goal of influencing the population. 

It can thus be interesting to consider the following problem: can a change in option quality ``compensate'' the impact of the presence of zealots in the decision outcome? Assume thus option $A$ to be better than option $B$, i.e., $Q_A>Q_B$, then we can find values of $\alpha \geq 1$ such that in absence of zealots the system converges to an equilibrium $a_{\mathrm{ini}}^* \lesssim 1$, namely the group achieve almost a total consensus for the option with the highest quality~\cite{CarlettiEtAl2023}. Then a fraction of zealots with opinion $B$, $0<\zeta<1$, is added to the population, and as shown above, this induces a shift in the system equilibrium: the asymptotic fraction of agents with opinion $A$ is smaller than without zealots, $a_{\mathrm{tmp}}^*<a_{\mathrm{ini}}^*\lesssim 1$. The question we are interested in is thus: would it be possible to increase the quality of opinion $A$, $Q_A^\prime>Q_A$, to compensate for the presence of zealots and restore a collective agreement to $a^*_{ini}$? 

To answer this question we decided to simplify the model by replacing the social network with an all-to-all coupling. This is not restrictive and allows us to focus on the main point; a similar conclusion can be drawn also for a generic social network. In this setting we have thus only three variables, $a(t)$ the fraction of agents with opinion $A$, $b(t)$ the fraction of susceptible agents with opinion $B$ and $\zeta$, the fraction of zealots committed to opinion $B$.

In absence of zealots, we have previously shown~\cite{CarlettiEtAl2023} that the equilibrium $a_{\mathrm{ini}}^*$ satisfies
\begin{equation*}
 P_\alpha\left(\frac{a_{\mathrm{ini}}^*}{a_{\mathrm{ini}}^*(1-Q)+Q}\right)=a_{\mathrm{ini}}^*\, .
\end{equation*}
The inclusion of zealots determines a new equilibrium, $a_{\mathrm{tmp}}^*$ solution of
\begin{equation*}
(1-\zeta) P_\alpha\left(\frac{a_{\mathrm{tmp}}^*}{a_{\mathrm{tmp}}^*(1-Q)+Q}\right)=a_{\mathrm{tmp}}^*\, .
\end{equation*}

In order to have $a_{\mathrm{ini}}^*$ as a solution of an interacting group that includes the same fraction $\zeta$ of zealots, we need to modify the quality of option $A$ and thus define a new quality ratio $Q^\prime=Q_B/Q_A^\prime$, to be able to solve
\begin{equation}
\label{eq:QA}
(1-\zeta) P_\alpha\left(\frac{a_{\mathrm{ini}}^*}{a_{\mathrm{ini}}^*(1-Q^\prime)+Q^\prime}\right)=a_{\mathrm{ini}}^*\, .
\end{equation}
Eq.~\eqref{eq:QA} can be solved with respect to $Q^\prime$ to obtain the following explicit expression
\begin{equation}
\label{eq:Qprime}
\begin{cases}
Q^\prime = \left[\frac{2}{\left(\frac{2a_{\mathrm{ini}}^*}{1-\zeta_B}-1\right)^\frac{1}{\alpha}+1}-1\right]\dfrac{a_{\mathrm{ini}}^*}{1-a_{\mathrm{ini}}^*}\quad \text{if }\dfrac{a_{\mathrm{ini}}^*}{a_{\mathrm{ini}}^*(1-Q)+Q}>\frac{1}{2}\\
Q^\prime = \left[\frac{2}{1-\left(1-\frac{2a_{\mathrm{ini}}^*}{1-\zeta_B}\right)^\frac{1}{\alpha}}-1\right]\dfrac{a_{\mathrm{ini}}^*}{1-a_{\mathrm{ini}}^*}\quad \text{if }\dfrac{a_{\mathrm{ini}}^*}{a_{\mathrm{ini}}^*(1-Q)+Q}<\frac{1}{2}\, . 
\end{cases}
\end{equation}

\begin{figure}[htp!]
\vspace{-2.0cm}
\centering
\includegraphics[scale=0.45]{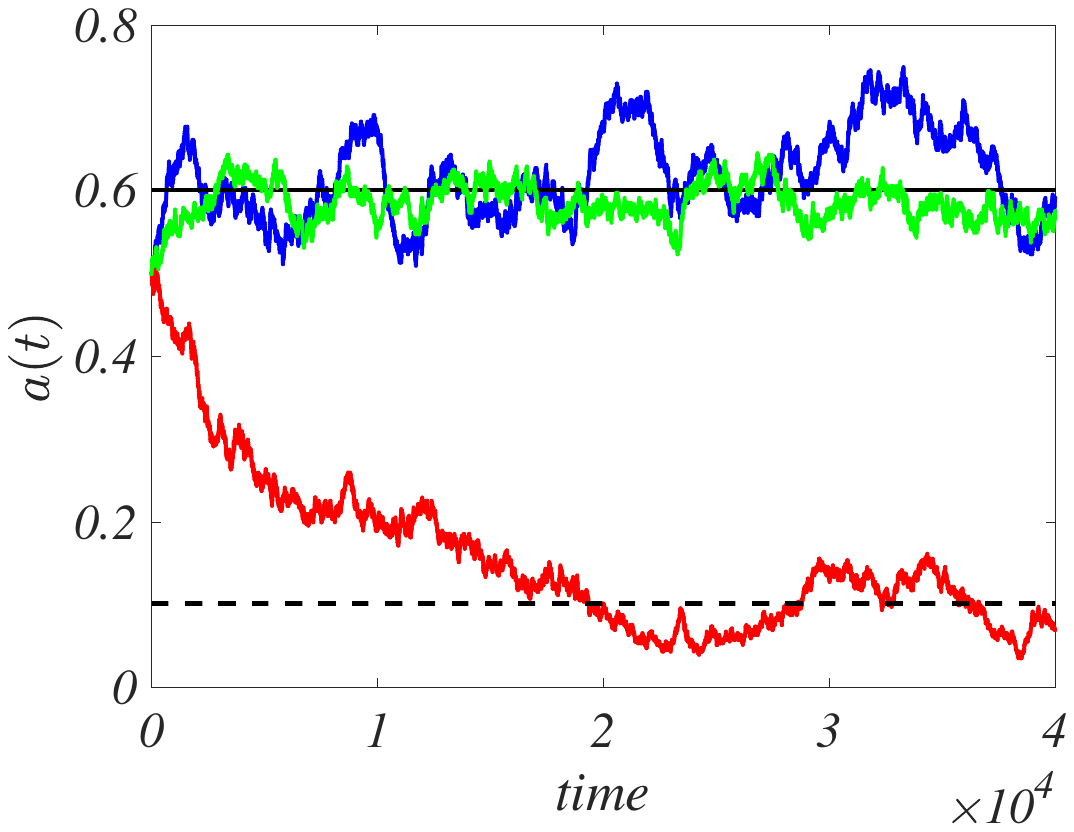}
\vspace{-3.5cm}
\caption{Numerical study of the relevance of quality versus zealots. We fixed $\alpha = 1.16$ and $Q_B = 0.9$, the blue line correspond to $Q_A=1$ and $\zeta=0$, the red line to $Q_A=1$ and $\zeta=0.2$ and the green line to $Q_A^\prime \sim 2.0756$ and $\zeta=0.2$.}
\label{fig:QvsZ}
\end{figure}

In Fig.~\ref{fig:QvsZ} we report the numerical results corresponding to the case $\alpha = 1.16$ for a generic set of parameters. The initial configuration is associated to $Q_A = 1$, $Q_B = 0.9$ and no zealots, $\zeta = 0$, resulting into the equilibrium $a_{\mathrm{ini}}^*\sim 0.602$ and one can observe that the numerical simulation (blue line) oscillates about such equilibrium (black horizontal line) because of the finite size effect induced by the discrete population. We then add a fraction $\zeta=0.2$ of zealots committed to $B$ and the system now converges (red line) to the new equilibrium $a_{\mathrm{tmp}}^*\sim 0.102$ (black dashed horizontal line). Finally, by using Eq.~\eqref{eq:Qprime}, we compute the value of $Q_A^\prime \sim 2.076$ which represent the new quality of option $A$, or simply the frequency with which opinions for $A$ are spread. By increasing option $A$'s quality to $Q_A^\prime$, the system returns to the original equilibrium $a_{\mathrm{ini}}^*$ (green line).

One can prove~\cite{CarlettiEtAl2023} that if $Q<1$ and $\alpha < 1$, in absence of zealots the system, intialized with half of agents committed to $A$ and half to $B$, will converge to a full consensus to $A$, namely $a_{\mathrm{ini}}^*=1$. In the previous section we have shown that in presence of zealots, the system cannot converge to a full $A$ consensus, hence we can never find a new value for the quality $Q_A$ able to return the equilibrium $a_{\mathrm{ini}}^*=1$ once zealots are present. To avoid this issue, we presented the above example by assuming $\alpha\geq 1$, values for which an equilibrium $a_{\mathrm{ini}}^* < 1$ can be achieved.

\section{Conclusions}
\label{cc}
In this paper, we presented the results of a study focused on a best-of-$n$ collective decision-making problem, with $n=2$ options of different quality. We analysed this problem through agent-based simulations and a mathematical model based on the heterogeneous mean-field approach.
In this study, the interactions among the individuals are defined using a social network whose nodes are agents, or decision-markers, and  edges the possible interactions among them. The original contribution of this paper is in highlighting the mutual effects existing between the following parameters of the model: i) the lower or higher heterogeneity of the scale-free network modelling the interactions between the agents; ii) the number of agents within the population that never change opinion (i.e., the zealots); iii) the pooling error which enables the control of agents' cognitive load; 
iv) the ratio in quality between the two options corresponding to the combination of the cost and benefit to each option, and v) the degree centrality (or the number of social connections) of the zealots. 
We have studied populations in which individuals are connected according to a scale-free network, and they select their opinion using decision mechanisms that differ in terms of their cognitive load, allowing to interpolate among existing models such as the voter model and the majority model. For each case, we have varied the number (or, in the mean-field model, the fraction) of zealots, all committed to the lowest quality option. The mathematical analysis of the combined effect of the considered parameters has been performed by determining the system equilibria and their stability.

The results have shown that the combined effect of these parameters has generated an articulated landscape characterised by different outcomes of the collective decision-making process. We have shown that, when susceptible agents employ opinion selection mechanisms characterised by pooling error $\alpha \sim 1$ (representing the voter model), both the number and the degree centrality of the zealots are elements that can induce the population to converge to the lowest quality option. In particular, the higher the number of zealots or the larger their degree centrality, the stronger their influence on the opinion dynamics. We have also shown that these effects are influenced by the nature of the opinion selection mechanisms employed by the susceptible agents. For example, the effect of the number of zealots on the opinion dynamics is not observed when susceptible agents make no pooling error, i.e., $\alpha = 0$ (representing the majority model). We have also shown that the connectivity structure modulates the influence of the zealots, indeed when the network becomes increasingly sparse and less heterogeneous, the effect of zealots is mitigated or largely reduced. 
In the future, we aim to extend to the proposed analytical model collective to best-of-$n$ decision-making by considering the concept of group interactions (also known as higher-order interactions\cite{battiston2021physics}) among the decision-makers and study how such higher-order can change the option dynamics.

\subsubsection*{Acknowledgements.} T.N. thanks the University of Namur for the financial support. A.R.~acknowledges support from the Belgian F.R.S.-FNRS, of which he was a Charg\'{e} de Recherches, and support from the Deutsche Forschungsgemeinschaft (DFG, German Research Foundation) under Germany's Excellence Strategy - EXC2117 - 422037984.

\subsection*{Declaration of competing interest}
The authors declare that they have no known competing financial interests or personal relationships that could have appeared
to influence the work reported in this paper.

\subsection*{Data availability}
No data was used for the research described in the article.

\bibliographystyle{plainnat} 
\bibliography{Refs}

    \end{document}